\documentclass[12pt,aps,amsmath,amssymb]{revtex4}
\usepackage{amsfonts}
\usepackage{bm}
\usepackage{graphicx}

\newcommand{\ga}{\gamma}
\newcommand{\om}{\omega}
\newcommand{\ep}{\varepsilon}
\newcommand{\dl}{\delta}
\newcommand{\bk}{{\bf k}}

\newcommand{\Tr}{{\rm Tr}\,}

\begin{document}
\title{Teaching the Environment to Control Quantum Systems}
\author{Alexander Pechen}\email{apechen@princeton.edu}
\author{Herschel Rabitz}\email{hrabitz@princeton.edu}
\affiliation{Department of Chemistry, Princeton
University, Princeton, New Jersey 08544, USA}

\begin{abstract}
A non-equilibrium, generally time-dependent, environment
whose form is deduced by optimal learning control is shown
to provide a means for incoherent manipulation of quantum
systems. Incoherent control by the environment (ICE) can
serve to steer a system from an initial state to a target
state, either mixed or in some cases pure, by exploiting
dissipative dynamics. Implementing ICE with either
incoherent radiation or a gas as the control is explicitly
considered, and the environmental control is characterized
by its distribution function. Simulated learning control
experiments are performed with simple illustrations to
find the shape of the optimal non-equilibrium distribution
function that best affects the posed dynamical objectives.
\end{abstract}

\date{\today}
\maketitle

\section{Introduction}
The manipulation of coherent atomic and molecular dynamics
often utilizes shaped electromagnetic fields as the
control. This topic is the focus of extensive theoretical
and experimental research~\cite{R0,R1,R2,R3,R4,R5,R6}
which relies on tailoring constructive and destructive
interferences between different dynamical pathways of a
quantum system. Many laboratory implementations of quantum
control with lasers use adaptive feedback learning
techniques~\cite{JR}.

In the present paper we consider the manipulation of
atomic and molecular dynamics with the control being a
tailored non-equilibrium, and generally time-dependent,
state of the surrounding environment. Different
non-equilibrium states of the environment can induce
correspondingly unique dynamical responses in the physical
system being controlled. Incoherent control by the
environment (ICE) is distinct from operations with
coherent control. Control by ICE affects a system through
dissipative dynamics and can be used to steer the system
from a pure or a mixed state into mixed and in some cases
pure states. Control by lasers normally affects the system
through Hamiltonian evolution and transforms pure states
into the pure states. In practice limitations will exist
on laser controls and the capabilities of ICE as well.
Thus, in the most general circumstance a shaped laser
pulse and a tailored non-equilibrium environment could be
combined into an overall tool to simultaneously best
perform control by Hamiltonian and dissipative dynamics.

This paper explicitly considers ICE implemented by two
types of environments: incoherent radiation (i.e., a gas
of photons) and gas of particles (i.e., electrons, atoms,
or molecules). The particles of an environment (i.e.,
photons or matter) are characterized by their momenta
$\bk$ and internal degrees of freedom indexed by $\alpha$.
For photons $\alpha$ denotes polarization while for atoms
or molecules it denotes their internal energy levels. The
environment is described by the distribution of mean
occupation numbers $n_{\bk,\alpha}(t)$ of its microscopic
states $|\bk,\alpha\rangle$ at time $t$. Thermal
equilibrium states are characterized by only a few
parameters, such as temperature and chemical potential,
which uniquely determine the corresponding equilibrium
distribution. Non-equilibrium states are characterized by
occupation numbers of all microscopic states and have much
richer structure. {\it The control considered with ICE is
the generally time dependent distribution function
$n_{\bk,\alpha}(t)$}. The shape of this function (i.e.,
its dependence on $\bk$, $\alpha$, and $t$) together with
the interaction Hamiltonian determines the dynamics of the
system under control. Although operation with ICE does not
induce coherent dynamics, nevertheless the shaping of
$n_{\bk,\alpha}(t)$ over $\bk,\alpha$, and $t$ provides
considerable flexibility for system manipulation. Many
control objectives can be expressed in terms of creating
specific mixed states of a system, which should be
amenable to utilizing non-equilibrium environmental states
for their preparation. Certain pure states can also be
reached with ICE. For example, incoherent radiation can
steer a three-level $\Lambda$-atom from the ground state
to the intermediate excited state, a gas can steer a
two-level atom through collisions to the excited state,
etc.

The practical creation of controls for implementing ICE is
important to consider. First, incoherent non-equilibrium
radiation as a control includes monochromatic incoherent
light, thermal radiation propagating in a medium with
frequency sensitive absorbtion (i.e., filters), and
radiation from luminescent emission. In the simplest case
monochromatic radiation can be either coherent or
incoherent, and in both circumstances it is composed of
waves with the same wavelength. In the second formulation
of ICE, a non-equilibrium gas, or more generally a
surrounding fluid or solid medium, of particles is
described by its distribution over momenta and internal
degrees of freedom (e.g., vibrational and rotational modes
of molecules or atomic energy levels). A non-equilibrium
state of a gas may be generated in a number of ways,
including through a sudden change in its thermodynamic
parameters. For example, if a lower energy level of a
molecule relaxes to equilibrium faster than an upper
level, then a sudden temperature drop can create a
non-equilibrium state with population inversion. A
non-equilibrium particle momentum distribution can be
created using selective excitation of the internal degrees
of freedom by a laser with subsequent collisional transfer
of the excitations into momentum modes. Lasers can have
dual roles in this general circumstance of (a) directly
addressing the system Hamiltonian for control as well as
(b) first addressing the environment for its subsequent
controlled manipulation of the system.

The main impediments to designing coherent optimal control
fields are the required detailed knowledge of the system
Hamiltonian and the need to solve the generally high
dimensional Schr\"odinger equation. The difficulties in
handling these two issues inevitably lead to significant
errors in the field designs which would likely result in
ineffective control in laboratory experiments. To overcome
these difficulties optimal learning control in the
laboratory is proving to be very successful~\cite{JR}. In
this fashion the system subjected to control is used as an
analog computer which solves its own Schr\"odinger
equation exactly in real time and with the true laser
field and system Hamiltonian. The capabilities of high
duty cycle laser pulse shaping along with fast observation
of the controlled outcome allows for efficient pattern
recognition algorithms (e.g., genetic
algorithms~\cite{GA}) to guide a sequence of experiments
to home in on the specified system control objective. The
same logic is proposed for ICE to find an optimal
non-equilibrium environment as a control, either alone or
possibly in conjunction with determining a coherent
control field.

This paper considers the following control problem: for a
given target state of the system $\rho^{\rm T}$, find a
distribution function of the environment
$n_{\bk,\alpha}(t)$ such that the corresponding induced
non-unitary system dynamics is steered from some initial
state $\rho_{\rm I}$ as close as possible to the state
$\rho^{\rm T}$. The corresponding objective function can
be chosen as
$J[n_{\bk,\alpha}(t)]=[\sum_{nm}(\rho_{nm}-\rho_{nm}^{\rm
T})^2]^{1/2}$, where $\rho_{nm}$ is an element of the
density matrix $\rho=\rho(t_f)$ at some final time $t_f$
evolved from $\rho_{\rm I}$ under the action of the
environment with distribution function
$n_{\bk,\alpha}(t)$. The problem of creating an effective
control based on significant environmental interactions
generally requires an adaptive learning control procedure.
The advantage of learning control lies in its ability to
find an optimal distribution $n_{\bk,\alpha}(t)$ even if
the details of the system, environment and their
interactions are not known. In the laboratory, practical
feedback signals would be of the form $\Tr\rho\Theta$,
where $\Theta$ is an observable operator, and the goal
would be to steer $\Tr\rho\Theta$ towards the desired
value $\Tr\rho^{\rm T}\Theta$. In the more general case,
expectation values of several possibly noncommutative
observable operators $\Theta_i$ could be used for
feedback, and the goal would be to steer the expectation
value of each $\Theta_i$ towards its desired value. Other
observable properties of the system, the environment, or
the means of generating its non-equilibrium state also can
be incorporated into the objective function.

Learning control driven by a GA is employed in the present
paper in simulations of the potential effectiveness of ICE
using either incoherent radiation or a gas as controls. In
these cases there is a Markovian regime (i.e., weak
coupling and low density) for the reduced dynamics of the
controlled system~\cite{SL,Sp,ldl,apv}. Master equations
with appropriate dissipative terms are reasonable models
in this regime. These equations are used here to simulate
evolution of the system's density matrix towards the
target under the influence of a non-equilibrium
environment. The general ICE concept is not limited to the
models used here, and in the laboratory fully
non-Markovian dynamics naturally would be included as
required.

An environment prepared impulsively in a non-equilibrium
state will evolve to equilibrium. In the simulations below
we consider control by stationary non-equilibrium
environments. In practice this requires keeping the
environment in a non-equilibrium state for a time which is
sufficiently long to perform the control. In the
simulations the master equations will be followed
temporally until a steady state is reached for the control
outcome. A stationary non-equilibrium state for incoherent
radiation can easily be maintained using steady sources.
In addition, a non-equilibrium state of a stationary gas
can be maintained by controlling the gas with a suitable
external field. An example of such control is preparation
of population inversion in the He--Ne gas-discharge laser.
In this case we may consider the gas of He atoms as the
control environment and the Ne atoms as the controlled
system. An electric discharge is passed through the He--Ne
gas to bring the He atoms into a non-equilibrium state.
Then He--Ne collisions transfer the energy of the
non-equilibrium state of the He atoms into the high energy
levels of the Ne atoms. This process creates a population
inversion in the Ne atoms and subsequent lasing. A steady
electric discharge can be used to keep the gas of helium
atoms in a non-equilibrium state to produce a CW He--Ne
laser. In the present simulations the actual nature of the
external control creating the non-equilibrium steady
distribution $n_\bk$ will not be explicitly described.
Rather in keeping with exploring the ICE concept, $n_\bk$
will be treated directly as the control for optimization
as a function of $\bk$. In the laboratory the external
control settings would be guided by the learning algorithm
using the system observations as a feedback signal.

Control by several lasers with incoherent relative phases
was considered in~\cite{cbs}. In this approach the
interference between different channels is manipulated by
changing the relative frequencies and intensities of the
lasers. Although the relative phases between the lasers
are incoherent, each of the lasers is a source of coherent
radiation which affects the system through induced
Hamiltonian dynamics. The present paper, in its part
devoted to control by radiation, uses as the control
totally incoherent radiation which affects the system
through induced dissipative evolution.

There is a relation between ICE and recently proposed
measurement-assisted~\cite{GR} and dual material-photonic
reagent control~\cite{MPC}. In all these situations
control is implemented, at least partially, through
induced dissipative dynamics. The ICE proposal directly
exploits the influence of an optimally shaped
environmental distribution function upon the dynamics of
the controlled system. Measurement-assisted control
exploits the back action of decoherence induced by
measurements. Material-photonic reagent control utilizes
dynamics driven by a laser as well as identification of an
optimal material for the system and/or the environment.
All these approaches to molecular or condensed phase
system manipulation can be considered as different
branches of the general perspective of introducing control
through same aspects of dissipative dynamics.

Sections~II and~III present simple illustrations of ICE using
incoherent radiation and a gaseous medium, respectively. The
formulations will be developed in the general context of a
coherent field being present for directly addressing the system
along with a time-dependent environmental control distribution
function $n_\bk(t)$. The test simulations will be performed with
steady distributions and without a coherent control field to serve
as a simple demonstration of the basic closed-loop ICE concept.
Even richer control should be possible with temporal control
densities operating along with a coherent field $\ep(t)$. Brief
concluding remarks are given in Section~IV.

\section{Incoherent radiation serving as a control}
This section considers control by non-equilibrium incoherent
radiation described by a distribution $n_\bk$ of the photon
momenta. In general, the polarization dependence of the incoherent
radiation also can be exploited as an additional control along
with the propagation direction in cases where spatial anisotropy
is important (e.g., a system consisting of oriented molecules
bound to a surface).

\begin{figure}[h]
\includegraphics[scale=0.45]{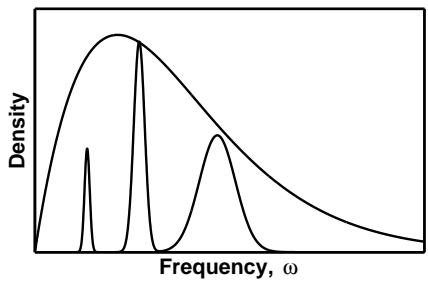}\caption{\label{fig1}
The Planck density of black body radiation (upper curve)
and the density (un-normalized) of non-equilibrium
radiation composed of three nearly monochromatic sources
(lower curve).}
\end{figure}

The thermal (i.e., equilibrium) distribution of photons at
temperature $T$ and frequency $\om$ is determined by
Planck's formula $N_\om=\om^2/\pi^2c^3
[\exp\left(\hbar\om/k_{\rm B}T\right)-1]$, where $c$ is
the speed of light, $k_{\rm B}$ and $\hbar$ are the
Boltzmann and the Planck constants (we set them to $1$ in
the sequel). Non-equilibrium incoherent radiation may have
an arbitrary distribution. Fig.~\ref{fig1} gives an
example of a thermal distribution and a non-equilibrium
distribution, which corresponds to incoherent radiation
composed of three nearly monochromatic components of
differing intensity. The latter distribution may be
produced either by filtering the thermal radiation or by
using independent monochromatic sources.\\

The master equation for a system simultaneously
interacting with a coherent electromagnetic field $\ep(t)$
and an environment with distribution $n_\bk(t)$ generally
has the form:
\begin{equation}\label{eq1}
\frac{d\rho(t)}{dt}=-i[H_0+H_{\rm
eff}-\mu\ep(t),\rho(t)]+{\cal L}_{\rm
diss}[n_\bk(t);\rho(t)].
\end{equation}
The coherent dynamics is generated by the system
Hamiltonian $H_0=\sum_n\ep_n P_n$ with eigenvalues $\ep_n$
and the corresponding projectors $P_n$, the effective
Hamiltonian $H_{\rm eff}$ originating from
system-environment interaction, dipole moment $\mu$, and
electromagnetic field $\ep(t)$. The effective Hamiltonian
commutes with $H_0$ and is not used in the following
simulations.

The dissipative dynamics generated by the incoherent
radiation distribution function $n_{\bk}(t)$ has the form
${\cal L}_{\rm diss}[n_\bk(t);\rho]={\cal L}^{^{\rm
Rad}}_{\rm diss}[n_\bk(t);\rho]=
\sum\limits_\om[\ga^+_\om(t)+\ga^-_{-\om}(t)]
(2\mu_\om\rho\mu^+_\om-\mu^+_\om\mu_\om\rho-\rho\mu^+_\om\mu_\om)$
where the sum is taken over all system transition
frequencies. Here the coefficients $\ga^\pm_\om(t)=\pi\int
d\bk\delta(|\bk|-\om)|g(\bk)|^2[n_\bk(t)+(1\pm1)/2]$
determine the transition rates between energy levels with
transition frequency $\om$ and
$\mu_\om=\sum\limits_{\ep_n-\ep_m=\om}P_m\mu P_n$. The
transition rates are the functions of the photon density
$n_\bk(t)$ and matrix elements of the dipole moment $\mu$.
The form-factor $g(\bk)$ determines the coupling of the
system to the $\bk$-th mode of the field. If for all
$\bk$, $n_\bk\equiv 0$, (i.e., the quantum system is in a
vacuum), then $\ga^-_\om\equiv 0$, and the coefficients
$\ga^+_\om$ together with dipole moment $\mu$ determine
the inverse lifetime of the system's energy levels. The
positivity of $\ga^\pm_\om(t)$ guarantees that the
off-diagonal elements of the density matrix vanish at
sufficiently long time.

As a specific example, numerical simulations are performed
for control of a four-level system by means of incoherent
radiation with the distribution $n_k$ in terms of the
magnitude of the photon momenta $k\equiv |\bk|$; no direct
coherent control is present such that $\ep(t)=0$. The
system has the free Hamiltonian $H_0={\rm diag}\{0;\,
11;\, 13;\,  24\}$ and dipole moment matrix
\begin{equation}\label{mu}
\mu= \left(\begin{array}{c c c c}
          0   & 0.8 & 0.3 & 0.5\\
          0.8 &   0 & 0.2 & 0.7\\
          0.3 & 0.2 &   0 & 1\\
          0.5 & 0.7 &   1 & 0
\end{array}\right)
\end{equation}
All transitions amongst the energy levels are allowed, and
the system is assumed to initially be in its ground state.
The goal of the control effort is to steer the system to a
target mixed state $\rho^{\rm T}$ (we consider several
examples for $\rho^{\rm T}$). The corresponding objective
function is chosen to have the form
$J[n_k]=[\sum_{nm}(\rho_{nm}-\rho^{\rm T}_{nm})^2]^{1/2}$
where $\rho_{nm}$ and $\rho^{\rm T}_{nm}$ are elements of
the system's density matrix and the target density matrix,
respectively. The goal is to minimize the objective
function at a sufficiently long time such that $\rho$ is
stationary.

\begin{figure}[t]
\begin{center}
\includegraphics[scale=1.1]{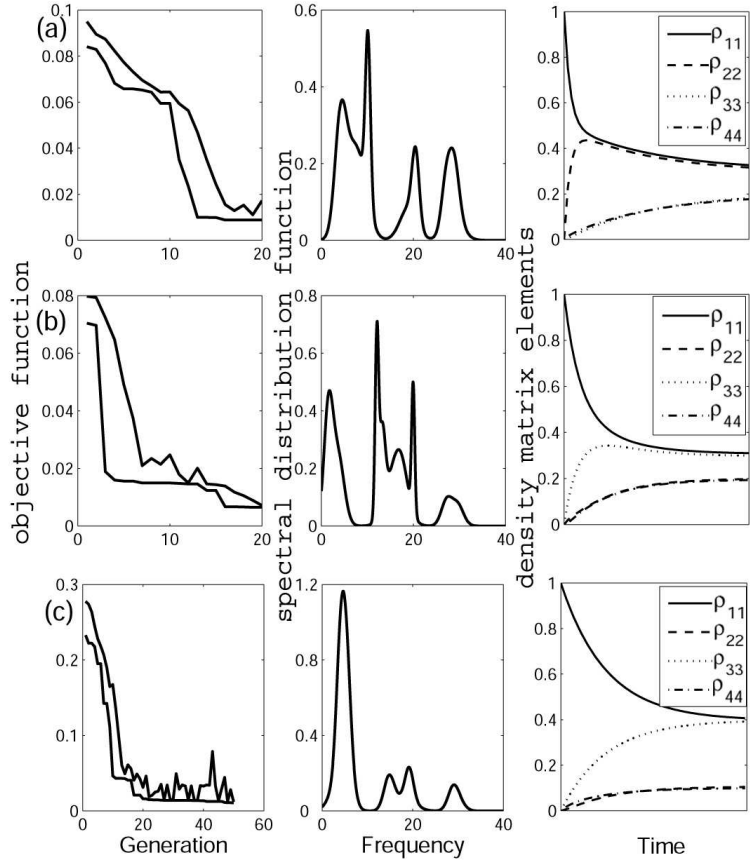}
\caption{\label{fig2} Results of ICE simulations with
tailored incoherent radiation as the control for target
states (a) $\rho^{\rm T}={\rm
diag}(0.3;\,0.3;\,0.2;\,0.2)$, (b) $\rho^{\rm T}={\rm
diag}(0.3;\,0.2;\,0.3;\,0.2)$, and (c) $\rho^{\rm T}={\rm
diag}(0.4;\,0.1;\,0.4;\,0.1)$. Each case shows: the
objective function vs GA generation, the optimal spectral
distribution vs frequency, and the evolution of the
diagonal matrix elements of the density matrix for the
optimal distribution. In the plots for the objective
function the upper curve is the average value for the
objective function and the lower one is the best value in
each generation.}
\end{center}
\end{figure}

Each distribution function $n_\bk$ determines the quantum
dynamical semigroup $P^t_{n_\bk}=e^{t{\cal L}_{n_\bk}}$,
$t\ge 0$ with generator ${\cal
L}_{n_\bk}[\,\cdot\,]:=-i[H_0,\cdot]+{\cal L}^{^{\rm
Rad}}_{\rm diss}[n_\bk;\cdot]$. An invariant state of the
semigroup $\rho_{\rm inv}$ is defined by ${\cal
L}_{n_\bk}[\rho_{\rm inv}]=0$ (thus $P^t_{n_\bk}(\rho_{\rm
inv})=\rho_{\rm inv}$). The solution of Eq.~(\ref{eq1})
with initial condition $\rho(t=0)=\rho_0$ has the form
$\rho(t)=P^t_{n_\bk}[\rho_0]$. If the coefficients
$\ga^\pm_\om$ are nonzero then the system density matrix
$\rho(t)$ will converge at long time to $\rho_{\rm inv}$.
For a given distribution function $n_\bk$ one can compute
the corresponding Lindblad operator ${\cal L}_{n_\bk}$ and
find its invariant. Here we consider the inverse problem:
given a target state $\rho^{\rm T}$ find a distribution
function of the environment $n_\bk$ which generates the
Lindblad operator ${\cal L}_{n_\bk}$ whose invariant state
is as close as possible to $\rho^{\rm T}$. In the case of
control by radiation only values of $n_\bk$ at the system
transition frequencies generally have an effect on the
dynamics. In principle, these values of $n_\bk$ could be
calculated if there was full knowledge of the system,
environment and their interaction. This circumstance is
rarely the case, and for a gaseous medium the situation is
even more complex, since generally all modes of the gas
contribute to the dynamics [see Eq.~(\ref{D2})]. Even with
all of these uncertain conditions learning control can be
effective because it only relies on laboratory
control-response observations.

The non-equilibrium radiation distribution function is
modelled as $n_k=\exp(-\beta k)
\sum_{i=1}^{10}\exp[-(k-k_i)^2/2D_i]/\sqrt{2\pi D_i}$ with
$\beta=1/20$. The parameters $D_i$ and $k_i$ are optimized
over ranges large enough to include all of the system
transition frequencies. A set of these parameters forms an
individual in the population for the GA to operate with.
Each individual determines a distribution function $n_k$
which is used to drive the evolution of the system density
matrix $\rho(t)$ towards its stationary form as
$t\to\infty$ to ultimately determine the objective
function. In practice, the time $t$ only needs to be taken
out to some small multiple of the longest timescale of the
system transitions.

A GA with crossover and mutation operators is used to find
the optimal values for $k_i$ and $D_i$. The number of
individuals in each population is $14$. The total of $20$
variables form an individual corresponding to ten
parameters $k_i$ and $D_i$. Each variable is coded into a
string of $20$ bits. The two most fit individuals are
always retained in the successive generation. The
remaining twelve individuals are produced from the parent
population using the crossover and mutation operators. The
fitness function determines the number of times each
individual from the parent generation is chosen to produce
offsprings in the next generation. The probability of
crossover is $P_x=0.9$ and of mutation (a bit flip) is
$P_m=0.7/L_{\rm ind}$, where $L_{\rm ind}=400$ is the
length of the bit string forming an individual.

The results of the simulations for three different target
states are presented in Fig.~\ref{fig2}. Each case
contains plots of the objective function vs generation,
the optimal spectral distribution function vs frequency,
and the corresponding evolution of the system density
matrix. In each case the target objective is met very well
with a suitable spectral distribution function. The
fitness function in cases (a) and (b) has small values for
the initial population because most randomly chosen
distribution functions $n_k$ induce states of the system
which are close to the equally populated state. Hence, the
search for a distribution function which steers the system
to a complex target state is a non-trivial problem. A
general expectation is that certain modes of radiation
will promote transitions to the target state whereas
others may be harmful for control. Thus, it is found that
distinct distributions of radiation energy are required to
most effectively steer the system into each particular
target state. Each of the radiation distribution functions
has components at all of the system transition
frequencies, and the mechanism of the control is not
simply evident in each case. Further analysis~\cite{abhra}
would be needed to identify the control mechanism. In
keeping with the logic of the adaptive control technique,
the learning algorithm deduces how to distribute the
radiation energy without specific knowledge or details of
the microscopic dynamics, including the Hamiltonian and
initial state of the system, as well as details of its
coupling to the non-equilibrium environment.
Fig.~\ref{fig2} shows that excellent results can be
achieved even with diverse choices for the target density
matrix.

Using incoherent radiation as the control clearly has some
limitations. For example, incoherent radiation can not
create population inversion in a two-level atom. However
incoherent radiation can steer a quantum system from a
pure state to a mixed one, possibly of a complex structure
as indicated in Fig.~\ref{fig2}, and in some cases
specific pure states can also be reached. Tailored
incoherent radiation also can be used jointly with a laser
field $\ep(t)$ to improve the degree of system control
when significant laser restrictions exist, such as bounds
on laser intensity or bandwidth.

\section{A gaseous medium serving as a control}
This section considers ICE with a non-equilibrium density
$n_\bk(t)$ of gas particles such as electrons, atoms or
molecules serving as the control. Quantum systems
interacting with such gases are described by master
equations whose dissipative generators are different from
${\cal L}^{^{\rm Rad}}_{\rm diss}$ in Section~II. The gas
is assumed to be sufficiently dilute such that the reduced
dynamics of the system is Markovian. In this case the
probability of simultaneous interaction of the system with
two or more particles of the gas is negligible and the
reduced dynamics is determined by two body scattering
events between one particle of the system and one particle
of the gas. The assumption of the rarity of the gas is not
a restriction for ICE, and dense gases might be used for
control as well.

The master equation for a system interacting with a
coherent electromagnetic field $\ep(t)$ and a gas has the
form of Eq.~(\ref{eq1}) with the dissipative generator
${\cal L}_{\rm diss}[n_\bk(t);\rho]={\cal L}^{^{\rm
Gas}}_{\rm diss}[n_\bk(t);\rho]$ specified by the
distribution function of the gas $n_\bk(t)$ and by the
$T$-operator (transition matrix) for scattering of the
system and a gas particle. A transition matrix element is
$T_{n,n'}(\bk,\bk')=\langle n,\bk|T|n',\bk'\rangle$, where
$|n,\bk\rangle\equiv|n\rangle|\bk\rangle$ denotes the
product state of the system discrete eigenstate
$|n\rangle$ (an eigenstate of the system's free
Hamiltonian $H_0$ with eigenvalue $\ep_n$) and a
translational state of the system and a gas particle with
relative momentum $\bk$. If the system is fixed in space
(we consider this case below corresponding to the system
particle being much more massive than the particles of the
surrounding gas) then $|\bk\rangle$ is a translation state
of a gas particle. The general case of relative system gas
particle motion can be considered as well using suitable
master equations. We assume that the particles of the gas
are characterized only by their momenta and do not have
internal degrees of freedom; otherwise, the state of one
particle of the gas should have the form
$|\bk,\alpha\rangle$, where $\alpha$ specifies the state
of the internal degrees of freedom. It is convenient to
introduce the notation $
T_\om(\bk,\bk'):=\sum\limits_{m,n:\,\,
\ep_m-\ep_n=\omega}T_{m,n}(\bk,\bk')|m\rangle\langle n|$.
The density of particles of the gas at momentum $\bk$ is
$n_\bk(t)$, the kinetic energy of a gas particle of mass
$M$ is $|\bk|^2/2M$, and $B$ is the set of transition
frequencies $\om$ of the system among the energy levels of
$H_0$. In this notation the dissipation generator is
\begin{eqnarray}
{\mathcal L}^{^{\rm Gas}}_{\rm
diss}[n_\bk(t);\rho]&=&2\pi\sum\limits_{\om\in B}\int d\bk
d\bk'\dl\left(\frac{|\bk'|^2}{2M}-\frac{|\bk|^2}{2M}+\om\right)n_\bk(t)
\Bigl[T_\om(\bk',\bk)\rho T^+_\om(\bk',\bk)\nonumber\\
&&-\frac{1}{2}\Bigl(T^+_\om(\bk',\bk)T_\om(\bk',\bk)\rho+\rho
T^+_\om(\bk',\bk)T_\om(\bk',\bk)\Bigr)\Bigr]\label{D2}
\end{eqnarray}
If the gas is at equilibrium with inverse temperature
$\beta$, then the density is stationary and has the
Boltzmann form $n_\bk(t)\equiv
n_\bk=C(\beta,n)\exp(-\beta|\bk|^2/2M)$, where the
normalization constant $C(\beta,n)$ is determined by the
condition $\int d\bk n_\bk=n$, where $n$ is the total
density of the gas. The structure of Eq.~(\ref{D2}) for
equilibrium gases has been discussed previously
in~\cite{ldl,apv} and for non-equilibrium stationary gases
in~\cite{p}. Non-equilibrium gases may be characterized by
generally time dependent distributions.
Equation~(\ref{eq1}) with ${\cal L}_{\rm
diss}[n_\bk(t);\rho(t)]={\cal L}^{^{\rm Gas}}_{\rm
diss}[n_\bk(t);\rho(t)]$ is the general formulation for
control by both a coherent electromagnetic field $\ep(t)$
and a non-equilibrium gas density $n_\bk(t)$. As a simple
illustration of ICE we only consider a simulation for
control by a static non-equilibrium distribution $n_\bk$.

Let $E_{i,n}=|\bk|^2/2M+\ep_n$ be the initial energy of
the total system consisting of one gas particle and the
controlled system before collision and let
$E_{f,m}=|\bk'|^2/2M+\ep_m$ be the final energy after the
collision. If for each transition frequency $\om$ there
are only two system levels $n$ and $m$ such that
$\om=\ep_n-\ep_m$, then Eq.~(\ref{eq1}) for the diagonal
elements of the density matrix reduces to the Pauli master
equation
\[
\frac{d\rho_{ll}(t)}{dt}=2\sum\limits_n
\bigl[w_{ln}\rho_{nn}(t)-w_{nl}\rho_{ll}(t)\bigr]
\]
where the transition probability $w_{mn}=\pi\int d\bk
n_\bk(t)\int
d\bk'\dl(E_{f,m}-E_{i,n})|T_{mn}(\bk',\bk)|^2$ between
levels $n$ and $m$ is explicitly determined by the
distribution function $n_\bk(t)$.

The quantity $\int
d\bk'\dl(E_{f,m}-E_{i,n})|T_{mn}(\bk',\bk)|^2$ defines the
scattering cross section between the system and one gas
particle. There are two possible strategies for
investigating the prospects for control by a
non-equilibrium gas. First, one may start with the
microscopic interaction Hamiltonian between the system and
a particle of the gas to compute the $T$-matrix for use in
the dissipative generator~(\ref{D2}). The second strategy
is to use experimentally measured cross sections for the
same purpose. As our purpose here is to illustrate the
prospect of closed-loop laboratory learning control with
ICE, we will simply use the first option in a model.
\begin{figure}[t]
\begin{center}
\includegraphics[scale=1.1]{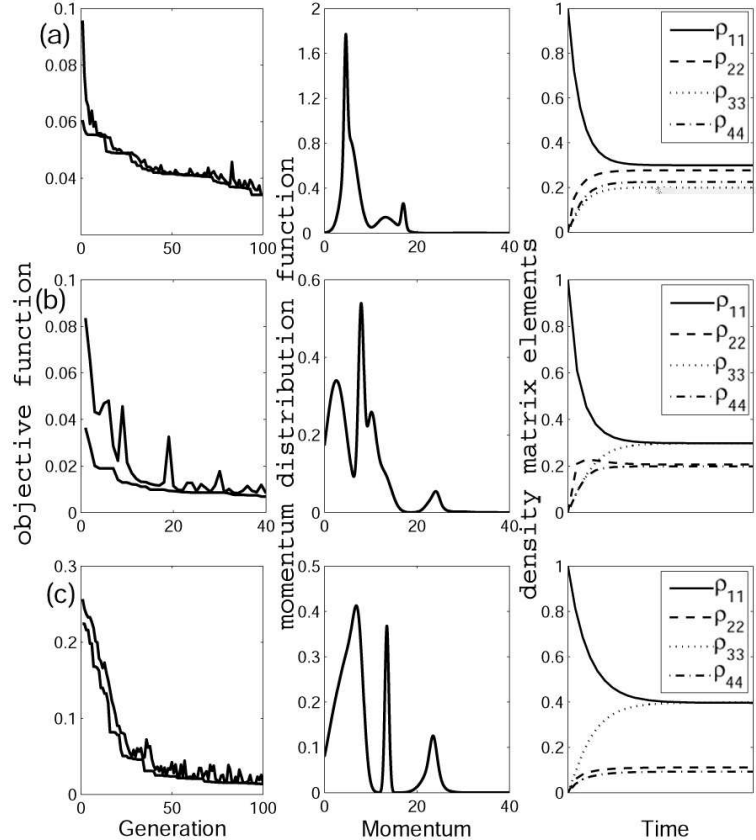}
\caption{\label{fig1ldl} Results of ICE simulations with a
surrounding non-equilibrium gas as the control for target
states (a) $\rho^{\rm T}={\rm
diag}(0.3;\,0.3;\,0.2;\,0.2)$, (b) $\rho^{\rm T}={\rm
diag}(0.3;\,0.2;\,0.3;\,0.2)$, and (c) $\rho^{\rm T}={\rm
diag}(0.4;\,0.1;\,0.4;\,0.1)$. Each case shows: the
objective function vs GA generation, the optimal
distribution vs momentum, and the evolution of the
diagonal elements of the density matrix for the optimal
distribution. In the plots for the objective function the
upper curve is the average value for the objective
function and the lower one is the best value in each
generation.}
\end{center}
\end{figure}

In the simulations we consider a four-level system with
the same free Hamiltonian $H_0$ in Section~II, immersed in
a dilute gas. The system is initially prepared in its
ground state and the goal of the control is to steer the
system to a target state $\rho^{\rm T}$ (we consider the
same target states and the same fitness function as in
Section~II). The interaction $V$ between the system and
one particle of the gas is considered to be weak with
matrix elements $V_{nm}(\bk,\bk')\equiv \langle
n,\bk|V|m,\bk'\rangle=\mu_{nm}g_n(\bk)g_m(\bk')$ defined
by the matrix $\mu$ of the form~(\ref{mu}) and by the
functions $g_n(\bk)$ chosen as characteristic functions of
the momentum magnitude $k\equiv|\bk|$,
$g_n(\bk)=\chi_{[a_n,b_n]}(k)$. Here $\chi_{[a_n,b_n]}(k)$
has unit value if $a_n\le k\le b_n$ and zero otherwise.
The parameters $a_n$ and $b_n$ are chosen randomly as
$a_1=2$, $b_1=12$, $a_2=9$, $b_2=24$, $a_3=3$, $b_3=17$,
$a_4=14$, and $b_4=26$. We chose $M=1$.

The matrix $\mu$ describes transitions between the
system's energy levels due to interaction with the gas and
the functions $g_n(\bk)$ describe the corresponding change
in the momentum of a gas particle. For a general $\mu$ its
diagonal elements are responsible for elastic scattering
of the system (these elements are zero in our case),
whereas the off-diagonal ones control the inelastic
events. Spontaneous emission from the upper levels is
assumed to be negligible, corresponding to the lifetimes
of the excited states being much longer than the inverse
transition rates due to collisions with the gas. Elastic
scattering and spontaneous emission can be included when
necessary, and all physical processes could naturally be
present in a laboratory closed loop experiment.

The weak nature of the interaction allows for replacing
the $T$-operator in Eq.~(\ref{D2}) by the interaction
Hamiltonian $V$. The control in the simulations is a
static distribution of the form $n_k=\exp(-\beta
k^2)\sum_{i=1}^{10}\exp[-(k-k_i)^2/2D_i]/\sqrt{2\pi D_i}$
with $\beta=0.01$. The parameters $k_i$ and $D_i$ are
optimally determined by the GA. This distribution together
with interaction Hamiltonian determines the dissipative
generator ${\cal L}_{\rm diss}^{^{\rm Gas}}$ according to
Eq.~(\ref{D2}) and the evolution of the system density
matrix according to the master equation~(\ref{eq1}) (with
$\ep(t)=0$ in this case). The goal of the control is to
find a stationary non-equilibrium state of the gas which
steers the system to a target state $\rho^{\rm T}$. If
desired, either the constraint of a fixed total energy of
the gas or its cost for minimizing its value could be
included by adding appropriate terms to the objective
function. Fig.~\ref{fig1ldl} gives the results of the
numerical simulation for different target states. The
simulations show that diverse mixed states can be reached
very well by manipulating the momentum distribution
function $n_k$ using ICE based on a learning algorithm.

\section{Conclusions} In this paper learning control with
a non-equilibrium environment is proposed as a means for
manipulating quantum systems. Two cases are simulated:
control by incoherent radiation and by a gas of particles.
The control is the distribution of mean occupation numbers
of the environment. The control affects the physical
system through tailored dissipative dynamics and allows
for steering an initial pure or a mixed state into a
complex target state. The search for an optimal control
distribution in ICE is performed by a learning control
strategy, which could be implemented in the laboratory
without detailed knowledge of the system Hamiltonian,
coupling to the environment, etc. The method can be
generalized by combining standard laser control and the
proposed ICE control. In the latter case a most
interesting situation would be attaining states which can
not be obtained by using either restricted lasers or a
non-equilibrium environment alone. An open issue is to
establish the degree of control, i.e., the set of states
reachable with ICE.

\begin{acknowledgments}
The authors acknowledge support from the National Science
Foundation and the Defense Advanced Research Projects
agency. A.P. acknowledges partial support from the grant
RFFI-05-01-00884-a.
\end{acknowledgments}


\begin{thebibliography}{99}
\bibitem{R0} D. Tannor and S. A. Rice, J. Chem. Phys. {\bf 83}, 5013
(1985).
\bibitem{R1} P. Brumer and M. Shapiro, Chem. Phys. Lett. {\bf 126}, 541 (1986).
\bibitem{R2} W. S. Warren, H. Rabitz and M. Dahleh, Science {\bf 259}, 1581 (1993).
\bibitem{R3} H. Rabitz, R. de Vivie-Riedle, M. Motzkus and K.~Kompa,
Science {\bf 288}, 824 (2000).
\bibitem{R4} S. A. Rice and M. Zhao, {\it Optical Control of Molecular Dynamics}
(Wiley: New York, 2000).
\bibitem{R5} M. Shapiro and P. Brumer, {\it Principles of the
Quantum Control of Molecular Processes}
(Wiley-Interscience: Hoboken, NJ, 2003).
\bibitem{R6} M. Dantus and V. V. Lozovoy, Chem. Rev. {\bf 104}, 1813 (2004).
\bibitem{JR} R. S. Judson and H. Rabitz, Phys. Rev. Lett. {\bf 68}, 1500 (1992).
\bibitem{GA} D. E. Goldberg, {\it Genetic Algorithms in Search, Optimization
and Machine Learning}
(Addison-Wesley, Reading, MA, 1989).
\bibitem{SL} H. Spohn and J. L. Lebowitz, Adv. Chem. Phys. {\bf 38}, 109
(1978).
\bibitem{Sp} H. Spohn, Rev. Mod. Phys. {\bf 53}, 569 (1980).
\bibitem{ldl} R. D\"umcke, Comm. Math. Phys. {\bf 97}, 331
(1985).
\bibitem{apv} L. Accardi, A. N. Pechen and I. V. Volovich,
Infin. Dimens. Anal. Quant. Probab. and Relat. Topics {\bf
6}, 431 (2003); {\it E-print:} math-ph/0206032.
\bibitem{cbs} Z. Chen, M. Shapiro and P. Brumer, Chem. Phys. Lett.
{\bf 228}, 289 (1994); Phys. Rev. A {\bf 52}, 2225 (1995).
\bibitem{GR} Gong J. and S. A. Rice, J. Chem. Phys. {\bf 120}, 9984 (2004).
\bibitem{MPC} R. Wu, A. N. Pechen and H. Rabitz, in preparation.
\bibitem{abhra} A. Mitra and H. Rabitz,
Phys. Rev. A {\bf 67}, 033407 (2003).
\bibitem{p} A. N. Pechen, in {\it QP-PQ: Quantum Probability and White Noise
Analysis}, Vol.~18, edited by M.~Sch\"urmann and U.~Franz
(World. Sci. Pub. Co., Singapore, 2005) pp. 428--447; {\it
E-print:} quant-ph/0607134.
\end{thebibliography}
\end{document}